\documentclass[epj]{svjour}

\usepackage{amssymb,amsmath}
\usepackage{graphics}
\usepackage{multirow}
\usepackage{dcolumn}
\usepackage{bm}
\usepackage{soul}
\usepackage{color}

\begin{document}
\title{Heavy-ion Collisions: Direct and indirect probes of the density and temperature dependence of E$_{sym}$}
\author{Z. Kohley\inst{1,2} \and S.~J. Yennello\inst{3,4}
}                     
\offprints{}          
\institute{National Superconducting Cyclotron Laboratory, Michigan State University, East Lansing, Michigan 48824, USA \and Department of Chemistry, Michigan State University, East Lansing, Michigan 48824, USA \and Cyclotron Institute, Texas A\&M University, College Station, Texas 77843, USA \and Chemistry Department, Texas A\&M University, College Station, Texas 77843, USA}
\date{Received: date / Revised version: date}
%
\abstract{
Heavy-ion collisions provide a versatile terrestrial probe of the nuclear equation of state through the formation of nuclear matter at a wide variety of temperatures, densities, and pressures.  Direct and indirect approaches for constraining the density dependence of the symmetry energy using heavy-ion collisions have been developed.  The direct approach relies on scaling methods which attempt to connect isotopic fragment distributions to the symmetry energy.  Using the indirect approach constraints on the equation of state are extracted from comparison of experimental results and theoretical transport calculations which utilize effective nucleon-nucleon interactions.  Besides exploring the density dependence of the equation of state, heavy-ion collisions are simultaneously probing different temperature gradients of nuclear matter allowing for the temperature dependence of the symmetry energy to be examined.  The current progress and open questions related to constraining the density and temperature dependence of the symmetry energy with heavy-ion collisions are discussed in the review.
} 

\authorrunning{Kohley and Yennello}
\titlerunning{Direct and indirect probes of the density and temperature dependence of E$_{sym}$}

\maketitle
\section{Introduction}
\label{intro}

\par
The nuclear equation of state (EoS) is most commonly discussed in reference to its density dependence.  The EoS is often expanded in terms of the asymmetry dependence as,
\begin{equation}\label{eos}
E(\rho,\delta) = E(\rho) + E_{sym}(\rho) \delta^{2}
\end{equation}
where the binding energy per nucleon of the infinite nuclear matter is shown as a function of density ($\rho$) and isospin concentration ($\delta$)~\cite{LI01,FUCH06,Tsang12}.  The isospin concentration or asymmetry is the difference in the proton and neutron densities, $\delta = (\rho_{n} - \rho_{p}) / (\rho_{n} +\rho_{p})$.  The first term of Eq.~\ref{eos} is not dependent on $\delta$ and represents the binding energy per nucleon of symmetric nuclear matter.  The coefficient of the second term, which has a $\delta^{2}$ dependence, is referred to as the symmetry energy.  The symmetry energy is the difference in energy between pure neutron matter ($\delta = 1$) and symmetric nuclear matter ($\delta = 0$).  A intensely studied topic of interest in the nuclear science community is the density dependence of the symmetry energy~\cite{LI01,Tsang12,Li08}.  The form of the density dependence of the symmetry energy is a property of asymmetric nuclear matter.  As discussed in a number of contributions to this special EPJA issue, the symmetry energy is a critical component for understanding neutron stars and the properties of finite nuclei~\cite{STEINER05,LATT04,CHEN05,LI02,KLAHN06,KRASTEV07,FURN02,Duc11,Roca08,Latt12,Stei12}.

\par
While the asymmetry sensitivity of the nuclear EoS is primarily discussed with reference to Eq.~\ref{eos}, a complete description of the EoS will define the relationship between all the state variables such as temperature, density, internal energy, and pressure.  For example, a significant amount of research has been devoted to examining the relationship between the thermal excitation energy ($E^{*}$) and temperature ($T$) of nuclei referred to as caloric curves~\cite{Poch95,Nat02,Bor08}.  Understanding the $E^{*} - T$ relationship was crucial in the development of the concept of a nuclear liquid-gas phase transition~\cite{Poch95,Sie83,Ell02}.  Since the symmetry energy is a component of the EoS, it too will have a dependence on other state variables such as the temperature.

\par
Most of the observables of finite nuclei that can be used to constrain the EoS provide a relatively limited range for accessing different regions of density, temperature, and pressure.  For example, giant monopole/dipole resonances~\cite{Li07,Tri08,Pie02}, pygmy resonances~\cite{Car10}, neutron skin thicknesses~\cite{Chen10,Abr12}, ground state binding energies~\cite{Moller12}, and isobaric analog states~\cite{Dan09} can all provide information on the nuclear EoS around the ground state of nuclear matter.  Heavy-ion collisions (HICs) provide experimentalists with an extremely versatile probe that can be used to produce nuclear matter at densities, temperatures, and pressures far away from that of stable nuclei.  By varying the energy and impact parameter of the collisions different regions of the nuclear EoS can be explored.  This aspect of HICs has been exploited for constraining the density dependence of the symmetry energy~~\cite{LI01,Tsang12,Li08,Shetty07,Koh10,BARAN05}.

\par
Extracting information on the symmetry energy from heavy-ion collisions can be accomplished through both direct and indirect methods.  In the direct approach, the measured fragment properties are used to extract information about the fragmenting system such as the density, temperature, or excitation energy.  Furthermore, information about the symmetry energy can be obtained directly from the fragment yields using methods such as isoscaling~\cite{Tsang01}.  Combining this information can provide a rich picture of the nuclear system and how its properties are related.  However, the assumptions associated with the scaling approaches can lead to significant uncertainties.  One of the ultimate achievements of the direct approach would be to measure the symmetry energy and density of a nuclear system simultaneously in an experiment allowing for $E_{sym}(\rho)$ to be extracted.  This was first accomplished by Natowitiz \emph{et al.} for nuclear matter at very low densities ($\rho < 0.009$~fm$^{-3}$)~\cite{Nat10,Kow07}.

\par
The other approach for extracting information about the nuclear EoS with heavy-ion collisions is through comparison of experimental results with theoretical predictions.  Transport models can simulate the dynamical evolution of the heavy-ion collisions which can be compared to the experimental data~\cite{Li08,BARAN02,ONO02,PAPA05}.  A key ingredient in the transport calculations is the effective nucleon-nucleon interaction or mean-field which defines the interaction between the nucleons.  Therefore, the interaction can be varied within the transport model in an effort to determine which form best reproduces a given set of experimental data.  Then information on the nuclear EoS can be inferred from the form of that nucleon-nucleon interaction.  For example, there is a large number of Skyrme interactions that can reproduce the binding energies of nuclei while the resulting density dependence of the symmetry energy can vary widely~\cite{Brown00}.  Using HICs the density dependence of the symmetry energy can be constrained from the interaction that best reproduces the experimental data.  The difficulty of this approach is contained in the details of the theoretical transport calculations which have many variables that can effect the HIC predictions~\cite{ONO06}.

\par
In the following review article, the recent results obtained through both the direct and indirect approaches will be discussed in relation to the density and temperature dependence of the symmetry energy.  The strengths and weaknesses of each approach are presented.  The results demonstrate that both approaches have, and will continue, to provide enlightening insights into the nuclear EoS.

\section{Direct Measurements}

\par
A direct measurement of the density, temperature, excitation energy, and symmetry energy of a nuclear system would provide a straightforward method of constraining the nuclear EoS.  In many regards, experimentalist have developed tools to allow for this type of information to be extracted directly from heavy-ion collsision measurements.  While a extensive amount of research has been devoted to the development and study of the different methods for measuring the temperature, excitation energy, and density of fragmenting nuclear systems~\cite{Bor08,Kow07,Viola04,Rad05,Wue10,Albergo85,Wada89,Hir84,Viola06}, these topics are not reviewed in this work as the focus remains on the symmetry energy.

\begin{figure}
\center
\resizebox{0.39\textwidth}{!}{\includegraphics{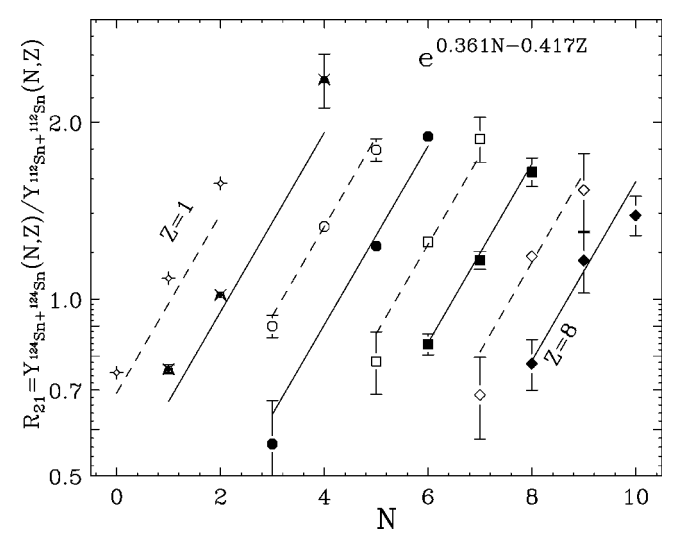}}
\caption{ Experimental yield ratios of $Z$~=~1 - 8 fragments from $^{112}$Sn~+~$^{112}$Sn and $^{124}$Sn~+~$^{124}$Sn heavy-ion collisions.  The solid and dashed lines represent the fits of the isoscaling relationship of Eq.~\ref{iso}.   Figure from Ref.~\cite{Tsang001} (Copyright 2001 by The American Physical Society).}
\label{f:Iso}
\end{figure}

\subsection{Isoscaling}

\par
The isotopic composition of the fragments resulting from a heavy-ion collision can provide a direct link to the symmetry energy~\cite{Col06,Ono03,Ono04}.  One approach to isolating the effect of the symmetry energy on the isotopic yields is the isoscaling relationship~\cite{Tsang01,Xu00}.  Isoscaling represents the exponential relationship exhibited in the isotopic yields of fragments between two systems with different neutron to proton ratios,
\begin{equation}\label{iso}
R_{21}(N,Z) = Y_{2}(N,Z) / Y_{1}(N,Z) = C \texttt{exp}(\alpha N + \beta Z)
\end{equation}
where the yield of a isotope with $N$ neutrons and $Z$ protons from system 1 (2) is shown as $Y_{1}(N,Z)$ ($Y_{2}(N,Z)$)~\cite{Tsang01,Tsang001,Xu00}.  The scaling parameters are $\alpha$ and $\beta$; $C$ is for normalization.  The convention is that system 2 has neutron-to-proton ratio ($N/Z$) larger than system 1.  An example of the isoscaling phenomena is shown in Fig.~\ref{f:Iso} from the experimental fragment yields of $^{112}$Sn~+~$^{112}$Sn and $^{124}$Sn~+~$^{124}$Sn heavy-ion collisions.

It has been shown that the symmetry energy can be linked to the isotopic yields through the relationship,
\begin{equation}\label{csym}
\alpha = \frac{4 C_{sym}}{T} \left[ \left( \frac{Z_{1}}{A_{1}} \right)^{2} - \left( \frac{Z_{2}}{A_{2}} \right)^{2} \right] = \frac{4 C_{sym}}{T} \Delta
\end{equation}
where $C_{sym}$ is the symmetry energy coefficient, $T$ is the temperature of the source, and the $(Z/A)^{2}$ terms are the source asymmetry from system 1 and 2~\cite{Tsang001,Bot02}.  The difference in the source asymmetry is often referred to as $\Delta$.    This relationship provides a powerful connection between the directly measured fragment yields and the symmetry energy.  From Eq.~\ref{csym} it is immediately clear that this relationship requires the two systems to have, ideally, the same temperature.  Theoretical examination has suggested that the fragment formation in heavy-ion collisions occurs at low-density and finite temperatures indicating that the isoscaling should be an appropriate tool to probe the symmetry energy~\cite{Ono04}.

It is also worth noting that the $C_{sym}$ of Eq.~\ref{csym} actually represents the symmetry free energy ($F_{sym}$) and is related to the internal symmetry energy through $E_{sym} = F_{sym} + TS_{sym}$, where $S_{sym}$ is the symmetry entropy~\cite{Nat10,Leh09,Ono04,Xu07}.  In most cases, the extracted $C_{sym}$ is referred to as the symmetry energy, which is likely a reasonable assumption around the saturation density~\cite{Kow07,Ono04}.  However, at sufficiently low densities the $TS_{sym}$ term is significant and, as shown by Natowitz \emph{et al.}, must be addressed~\cite{Nat10,Kow07,Wada12}.  Throughout Section 2 the symbol $C_{sym}$ is used, in comparison to $E_{sym}$, to represent values extracted from the isocaling method.

\subsection{$T$ and $\rho$ dependence of $C_{sym}$}

\begin{figure}
\center
\resizebox{0.37\textwidth}{!}{\includegraphics{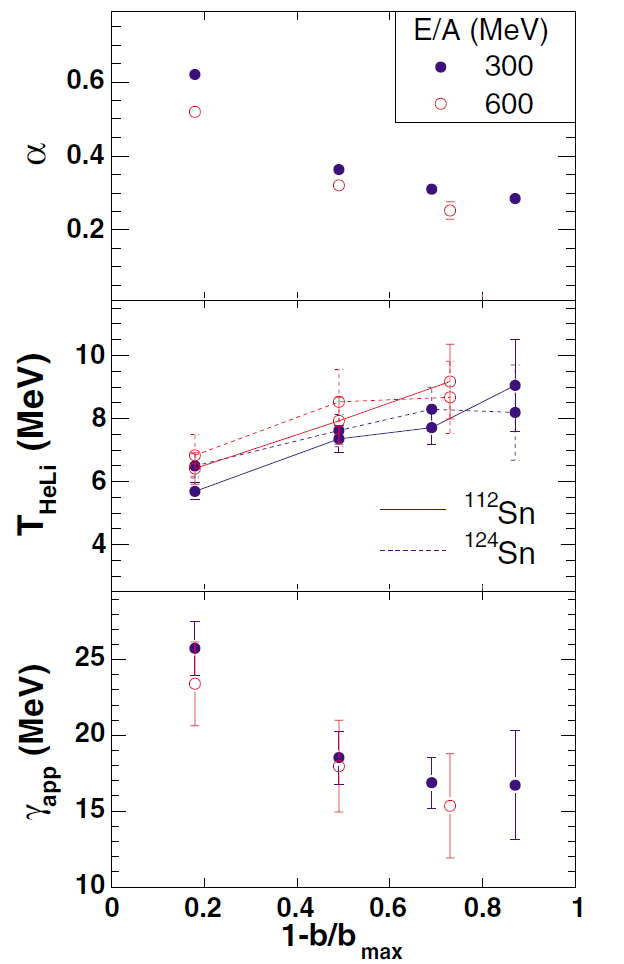}}
\caption{(Color online) Relationship between isoscaling parameter $\alpha$, temperature, and symmetry energy coefficient extracted from $^{124,112}$Sn~+~$^{12}$C reactions measured by the INDRA/ALADIN collaboration. Figure from Ref.~\cite{Lef05}  (Copyright 2005 by The American Physical Society).}
\label{f:indra}
\end{figure}

\par
An example of how the isoscaling method can be used to glean information about the symmetry energy is shown in Fig.~\ref{f:indra}.  Le F\`{e}vre \emph{et al.} used the isoscaling relationship to extract the $\alpha$ coefficient from the fragmentation of $^{124,112}$Sn~+~$^{12}$C as a function of the impact parameter at 300 and 600 MeV/A~\cite{Lef05}.  Furthermore, the temperature of the fragmenting system was directly calculated from the experimental data using the Albergo method~\cite{Albergo85}, as shown in the middle panel.  The $T$ between the $^{112}$Sn and $^{124}$Sn was assumed equivalent within the uncertainties and thus, Eq.~\ref{csym} was used to extract $C_{sym}$ (referred to as $\gamma_{app}$ in the bottom panel).

\par
\begin{figure}
\center
\resizebox{0.39\textwidth}{!}{\includegraphics{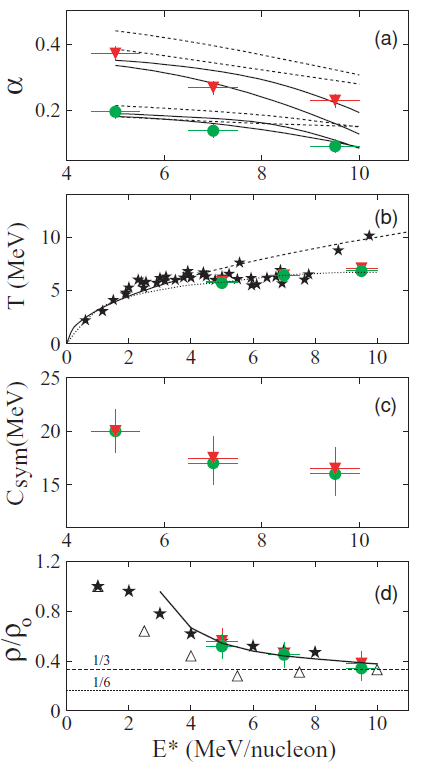}}
\caption{(Color online) The isoscaling parameter $\alpha$, temperature, symmetry energy coefficient, and density as a function of excitation energy was extracted from Fe + Fe and Ni + Ni systems (inverted triangles), and Fe + Ni and Ni + Ni systems (solid circles).   Figure from Ref.~\cite{Shetty07}  (Copyright 2007 by The American Physical Society).}
\label{f:shetty}
\end{figure}

From the results of Fig.~\ref{f:indra} it would be tempting to extract information about the relationship between $T$ and $C_{sym}$.  The results show an inverse relationship between $T$ and $C_{sym}$.  As $T$ rises from 6~MeV to 8~MeV a drastic drop in $C_{sym}$ occurs from about 25~MeV to 15~Mev.  However, it is important to recognize that in the heavy-ion collisions it is very difficult to vary a single property, such as $T$, while keeping the other properties of the system constant. The relatively small change in $T$ is accompanied by a large change in the centrality of the collisions, ranging from central ($1-b/b_{max} = 0$) to peripheral-like ($1-b/b_{max} = 0.2$) collisions.  The results of Shetty \emph{et al.}, shown in Fig.~\ref{f:shetty}, provide additional insight into the issue~\cite{Shetty07}.  Shetty \textit{et al.} measured the isotopic yields of fragments from Ni and Fe systems at energies of 30 to 47~MeV/A~\cite{Shetty07}.  The isoscaling parameter, temperature, and symmetry energy coefficient were all determined as a function of the excitation energy.  Similar to the results of Le F\`{e}vre \emph{et al.}, there is a slight increase in the temperature of the source which is correlated with a significant decrease in the symmetry energy.  Shetty \textit{et al.} also estimated the density of the system at break-up using the expanding Fermi-gas model and the result is shown in the bottom panel of Fig.~\ref{f:shetty}~\cite{Shetty07,Shetty_09}.  Over the examined excitation energy region, roughly 5~-~10~MeV/A, there is a significant drop in the density.  Thus, the decrease in the symmetry energy observed by Shetty and Le F\`{e}vre is likely a response to the decreasing breakup density of the system rather than the slight increase in temperature.  The corresponding density dependence of the symmetry extracted from the results of Shetty \textit{et al.} agree well with current constraints.

\par
\begin{figure}
\center
\resizebox{0.45\textwidth}{!}{\includegraphics{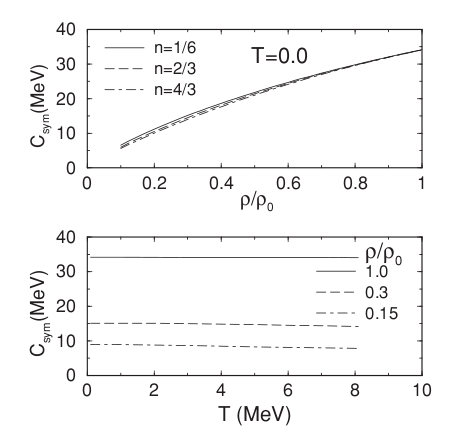}}
\caption{(Color online) Predicted density dependence of the symmetry energy with $T~=~0$ and temperature dependence of the symmetry energy at various densities.   Figure from Ref.~\cite{Sam07} (Copyright 2007 by The American Physical Society).}
\label{f:simT}
\end{figure}

The relatively weak $T$ dependence of the symmetry energy extracted from the experimental data in Fig.~\ref{f:shetty} is in agreement with most theoretical predictions~\cite{Sam07,Xu07,Li_06,Dean02,De12,Bra09,De_12,Beh09}.  However, a more detailed account of the theoretical approaches and results addressing the temperature dependence of the symmetry energy is presented in this special EPJA issue by Agrawal \emph{et al.}.   In general, the density dependence of the symmetry energy is expected to be stronger than the $T$ dependence, however this is dependent on the interaction and method used for the calculation.  An example is shown in Fig.~\ref{f:simT}, where the density and temperature dependencies of the symmetry energy were calculated using the Seyler-Blanchard interaction within the Thomas-Fermi formulation~\cite{Sam07}.  The results also show a weak dependence of the symmetry energy on $T$.  A difficult but important experimental measurement would be to try and isolate the $T$ dependence of the symmetry energy.  The $T$ dependence of the symmetry energy is particulary important for understanding the dynamics of the collapose and explosion of massive stars~\cite{Dean02,De_12,STEINER05,Jank07}.  In these stages the nuclear matter of the massive star is warm ($T > 0$~MeV).  Therefore, it is critical to understand the evolution of the symmetry energy with temperature.

\par
Accomplishing a direct measurement of the $T$ dependence of the symmetry energy would require identifying a reaction mechanism(s) which would produce sources with relatively constant densities but varying temperatures.  The results of Xu \textit{et al.} suggested that the largest $T$ dependence of the symmetry energy occurs when the breakup density of the system is very low~\cite{Xu07}.  Thus probing very low-density nuclear matter may provide the best oppurtunity to isolate the $T$ dependence from the density dependence.  Natowitz and collaborators have shown how the temperature, density, and symmetry energy can all be reliably extracted at low densities~\cite{Nat10,Wada12,Qin12,Rop13}.  While the results show a strong correlation between the density and temperature, these types of measurements may provide the best avenue forward in attempting to learn about the $T$ dependence of the symmetry energy.  Even if it is not possible to completely isolate the temperature dependence, simultaneous measurements of the temperature, density, and symmetry energy could significantly constrain the theoretical models.


\subsection{Definition of $\Delta$}

\par
\begin{figure}
\center
\resizebox{0.36\textwidth}{!}{\includegraphics{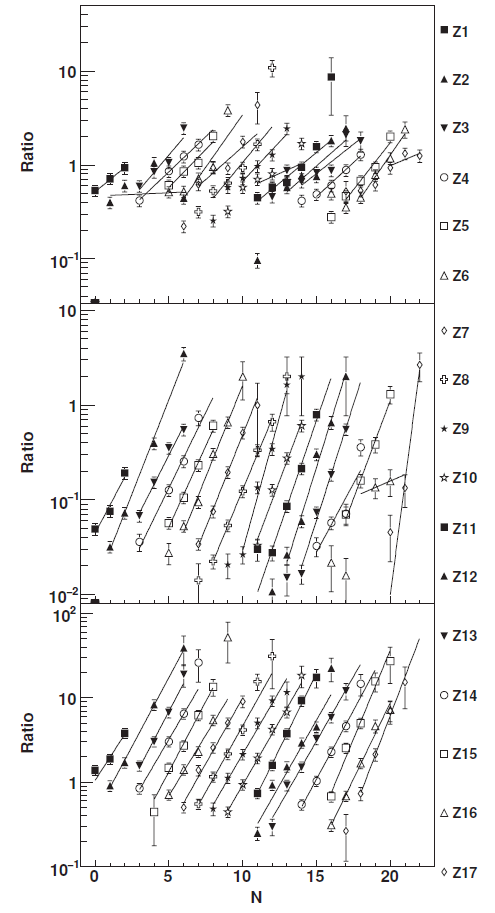}}
\caption{(Color online) Isotopic yield ratios from 35~MeV/A $^{86}$Kr~+~$^{64}$Ni and $^{78}$Kr~+~$^{58}$Ni systems.  The solid lines represent isoscaling fits following Eq.~\ref{iso}.  The difference between the isoscaling of the top, middle, and bottom panels is discussed in the text. Figure from Ref.~\cite{Wue09}  (Copyright 2009 by The American Physical Society).}
\label{f:wue}
\end{figure}

From the relationship between isoscaling and the symmetry energy (Eq.~\ref{csym}) it is clear that improvements in the experimental determination of $\alpha$, $\Delta$, and $T$ will aid in the extraction and investigation of $C_{sym}$.  Wuenschel \emph{et al.} showed that the quality of the fragment isoscaling could be significantly improved through accurately defining the $\Delta$ of the emitting sources~\cite{Wue09}.  In Fig.~\ref{f:wue} the isotopic yield ratios are shown from the 35~MeV/A $^{86}$Kr~+~$^{64}$Ni and $^{78}$Kr~+~$^{58}$Ni reactions using three different source definitions.  In the top panel, the source, and therefore $\Delta$, were defined simply by the $Z/A$ of the two reaction systems.  This procedure was originally adapted in early isoscaling studies.  The results of Wuenschel \emph{et al.} show that the fragment yields exhibit only a modest scaling.  Rather than assuming that the $N/Z$ of the source is constant for a given beam-target combination, realistically each collision will produce an emitting source with a different $N/Z$.  Therefore, primary quasi-projectile sources were reconstructed from the experimental data on an event-by-event basis.  Then two sets of events, each with a relatively narrow source (or quasi-projectile) $Z/A$, could be selected and the isotopic ratios examined.  The results of this procedure are shown in the middle panel of Fig.~\ref{f:wue} and demonstrate a marked improvement in the isoscaling.  The source $Z/A$ was further refined by including the measurement of free neutrons emitted in the reaction.  By selecting sources based on the free neutron corrected $Z/A$ of the quasi-projectile source, the isotopic yield ratios were shown to exhibit excellent isoscaling features (bottom panel).  Thus, the extracted $\alpha$ and the source $Z/A$, calculated event-by-event, are of much higher quality and should result in a more accurate determination of $C_{sym}$.  A similar effect was shown in the work of Galanopoulos \emph{et al.} for mass $\sim$40 quasi-projectiles~\cite{Gal10}.

\par
\begin{figure*}
\center
\resizebox{0.65\textwidth}{!}{\includegraphics{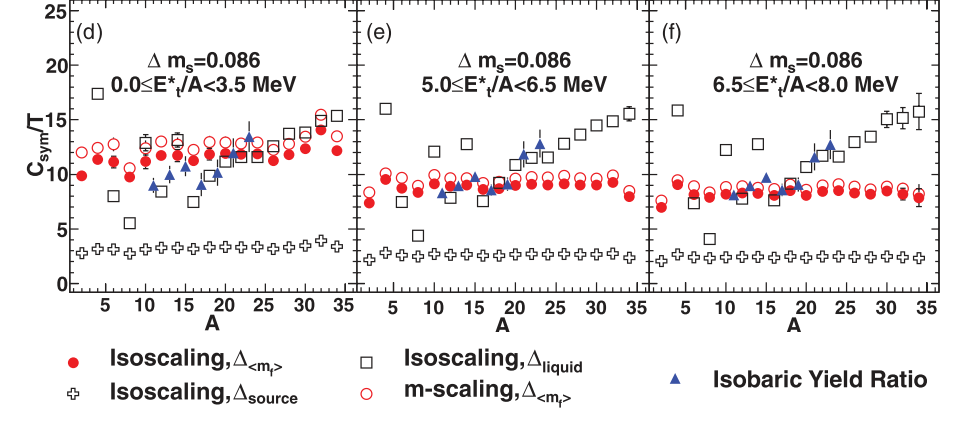}}
\caption{(Color online) Extracted $C_{sym}/T$ from the isoscaling, m-scaling, and isobaric yield ratios methods.  The isoscaling results are also shown using three different definitions of $\Delta$ ($\Delta_{<m_{f}>}, \Delta_{source}, \Delta_{liquid}$).  Figure from Ref.~\cite{Mar12}  (Copyright 2012 by The American Physical Society).}
\label{f:iso_compare}
\end{figure*}

While the results of Wuenschel \emph{et al.} showed the importance of accurately determining the $N/Z$ of the emitting source, different definitions of $\Delta$  have been proposed and can significantly alter the extracted $C_{sym}$ values~\cite{Ono03,Tsang001,Trip11}.  In Fig.~\ref{f:iso_compare} $C_{sym}/T$ was extracted as a function of the fragment mass from the isoscaling of the fragments emitted from the reconstructed quasi-projectiles of Zn~+~Zn and Ni~+~Ni reactions~\cite{Mar12}.  The open crosses were calculated using the ``standard'' definition of $\Delta$ defined in Eq.~\ref{csym}~\cite{Tsang001}, which is also the definition used in Figs.~\ref{f:indra}, \ref{f:shetty}, and \ref{f:wue}.  The closed red circles ($\Delta_{<m_{f}>}$) and open squares ($\Delta_{liquid}$) show the results of the same isoscaling analysis with different definitions of $\Delta$ (see Ref.~\cite{Mar12} for extended discussion).  In general the different definitions of $\Delta$ stem from different theoretical frameworks.  The ``standard'' definition of Eq.~\ref{csym} ($\Delta_{source}$) comes from the statistical multifragmentation and expanding emitting sources models~\cite{Bot02,Tsang001}.  The $\Delta_{liquid}$ definition derived from a study based on the dynamical Antisymmetrized Molecular Dyanmics model~\cite{Ono03,Ono04} and the $\Delta_{<m_{f}>}$ definition was derived on the basis of a Landau free energy approach~\cite{Trip11}. As shown, significant variations exist between all three definitions.  The $\Delta_{source}$ and $\Delta{<m_{f}>}$ definitions provide information on the average isospin of the sources whereas the $\Delta_{liquid}$ is a function of the $Z$ of the fragment and, thus, varies based on the average isospin of a given element produced in the reaction.  Further distinction of these differences is provided in the following sections as it also applies to the different scaling method approaches.


\subsection{Scaling Methods}

\par
Different scaling methods have been proposed to connect isotopic fragment yields with the symmetry energy coefficient.  The results from the m-scaling approach~\cite{Haung_10} and isobaric yield ratio method~\cite{Haung10} are shown in Fig.~\ref{f:iso_compare} along with the different definitions of $\Delta$ used in the isoscaling method.  Overall, the comparative study shows that the magnitude and $A$-dependence of the extracted $C_{sym}$ can vary significantly depending on the underlying scaling theory and the definition of $\Delta$.  A couple of observations can be extracted from Fig.~\ref{f:iso_compare}:
\begin{itemize}
  \item The isoscaling and m-scaling approaches provide similar results when the definition of $\Delta$ is the same (open and closed red circles).
  \item When the source is defined by it's average isospin ($\Delta_{source}$ or $\Delta_{<m_{f}>}$) the extracted $C_{sym}/T$ is relatively constant as a function of mass.
  \item $C_{sym}/T$ varies as a function of mass when using the isobaric yield ratio method or using the $\Delta_{liquid}$ definition.
  \item The ``standard'' isocaling approach with $\Delta_{source}$ produces significantly lower values of $C_{sym}/T$ in comparison to the other methods.
\end{itemize}

\par
The results of Fig.~\ref{f:iso_compare} present a complicated picture of the ability of scaling methods to provide insight in the symmetry energy.  Defining the correct approach is clearly a high priority and may require careful examination of the assumptions of each approach.  With a simple argument we can attempt to discern which approach provides the most realistic results.  Using the Fermi gas model relationship of $E^{*} = aT^{2}$, with the level density parameter $a = A/8$, an $E^{*} = 3.5$~MeV/A should correspond to $T\approx5$~MeV.  Taking $T = 5$~MeV, $C_{sym}$ can then be calculated from the measured $C_{sym}/T$ of Fig.~\ref{f:iso_compare}.  For the isoscaling method with $\Delta_{source}$ (open crosses) this leads to $C_{sym}\approx20$~MeV, while all the other approaches give $C_{sym}/T$ $\approx$ 10 corresponding to an erroneously large value of $C_{sym}$~=~50~MeV.  Further investigation is required to decisively determine the optimum approach for extracting information on $C_{sym}$ from scaling studies.

\subsection{Secondary Decay}

\begin{figure}
\center
\resizebox{0.48\textwidth}{!}{\includegraphics{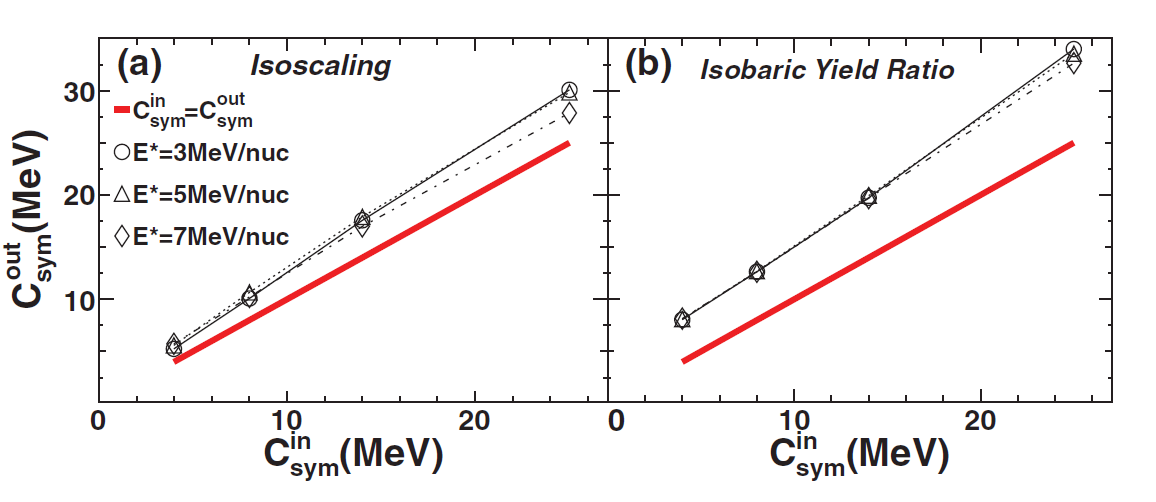}}
\caption{(Color online) The extracted symmetry energy coefficient ($C_{sym}^{out}$) from the primary fragment yields from the decay of hot $^{78,86}$Kr sources as a function of the input symmetry energy ($C_{sym}^{in}$) from the statistical multifragmentation model (SMM). The results from the isoscaling method using the ``standard'' source ($\Delta_{source}$) and the mircocanonical $T$ from the SMM calculation are shown in the left panel.  The isobaric yield ratio approach, also using the microcanonical $T$, is shown in the right panel.  The red solid lines provide references for $C_{sym}^{out} = C_{sym}^{in}$. Figure from Ref.~\cite{Mar13}  (Copyright 2012 by The American Physical Society).}
\label{f:smm}
\end{figure}

\par
Another issue that has been highly debated is the role secondary decay plays in the extraction of $C_{sym}$.  The theoretical connections between the isoscaling parameter and the symmetry energy (Eq.~\ref{csym}) is defined for the isotopic yields of the primary fragments emitted from a hot equilibrated source at a given temperature.  The hot primary fragments can then undergo further breakup referred to as secondary decay.  For example, the hot source may emit a $^{8}$Be fragment that will secondary decay into $2\alpha$, which would enhance the measured $\alpha$ yield beyond the primary $\alpha$ yield.  The effect of the secondary decay should be most significant for light and intermediate mass fragments in comparison to heavier residues (further discussion in Section~\ref{sec:heavy}).   Statistical multifragmentation models can simulate the decay of a hot source, including the secondary decay, where the fragment emission probability is a function of the binding energy.  Therefore, the symmetry energy coefficient can be varied within the model to examine how the isotopic fragment yields would change.  In general, previous works indicate that the secondary decay does modify the extracted isoscaling parameter~\cite{Mall13,Zhou11,Bot02,Rad_05}.  However, the extent of this effect is not well determined~\cite{Col06} and in some cases is predicted to be negligible~\cite{Tian08}.  Further work to provide a clear relationship between secondary decay and isoscaling would be highly valued.  Recently, Marini \textit{et al.} tested the secondary decay effects by varying the input symmetry energy coefficient ($C_{sym}^{in}$) in the SMM model and then calculating the symmetry energy coefficient using the isoscaling and isobaric yield ratio methods ($C_{sym}^{out}$)~\cite{Mar13}. As shown in Fig.~\ref{f:smm}, both methods provide a linear response of $C_{sym}^{out}$ as a function of $C_{sym}^{in}$ with the use of primary fragment yields.  While $C_{sym}^{in} \neq C_{sym}^{out}$, it is evident that the primary fragment yields can provide direct information about the symmetry energy.

\begin{figure}
\center
\resizebox{0.43\textwidth}{!}{\includegraphics{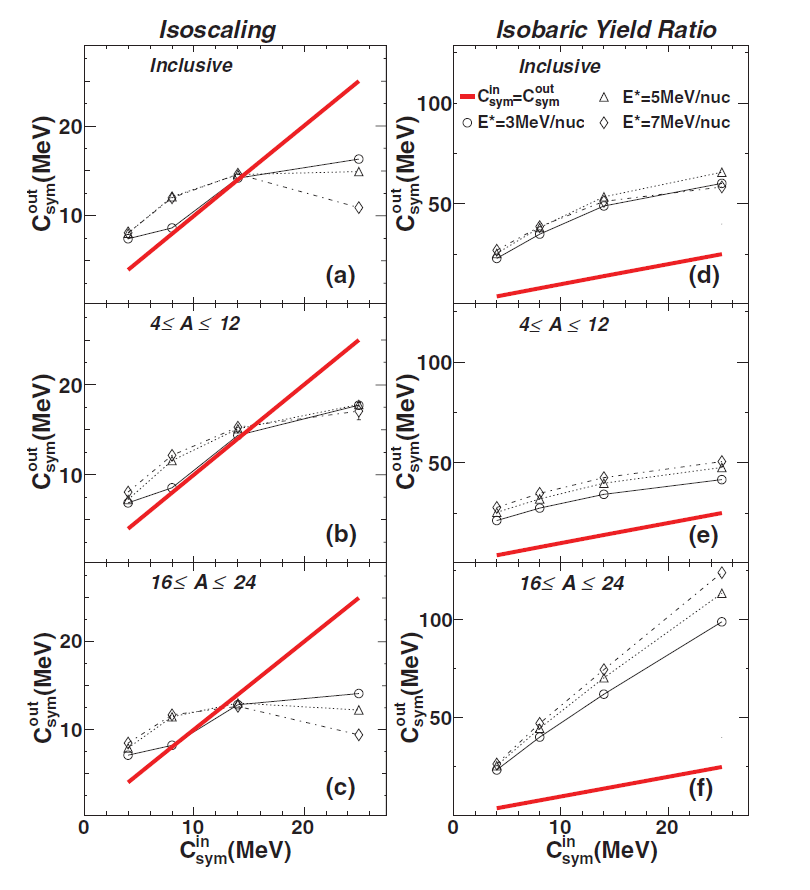}}
\caption{(Color online) Same as Fig.~\ref{f:smm}, except the secondary decay fragment yields were used in the analysis. Figure from Ref.~\cite{Mar13}  (Copyright 2012 by The American Physical Society).}
\label{f:smmcold}
\end{figure}

\par
The effect of the secondary decay can be observed in Fig.~\ref{f:smmcold}, which is from the same simulation shown in Fig.~\ref{f:smm} except the secondary fragment yields were used with the isoscaling and isobaric yield method to extract $C_{sym}^{out}$~\cite{Mar13}.  The different panels show the results using different ranges of fragments for the scaling analysis.  In general, the relatively strong correlation observed using the primary fragments with the isoscaling approach is lost due to the secondary decay.  In particular, at large values of $C_{sym}^{in}$ the extracted $C_{sym}^{out}$ can be hindered by 50$\%$.  Similar results have been observed in other calculations as shown in Refs.~\cite{Lef05,Ogul11}.  If the statistical multifragmentation models are accurately simulating the primary and secondary decays, then these results indicate the difficulty in directly extracting information on the symmetry energy from the cold fragment yields.  This is the reason why the bottom panel of Fig.~\ref{f:indra} was labeled as $\gamma$-apparent rather than $C_{sym}$.

\subsection{Heavy fragment isoscaling}
\label{sec:heavy}

\begin{figure}
\center
\resizebox{0.33\textwidth}{!}{\includegraphics{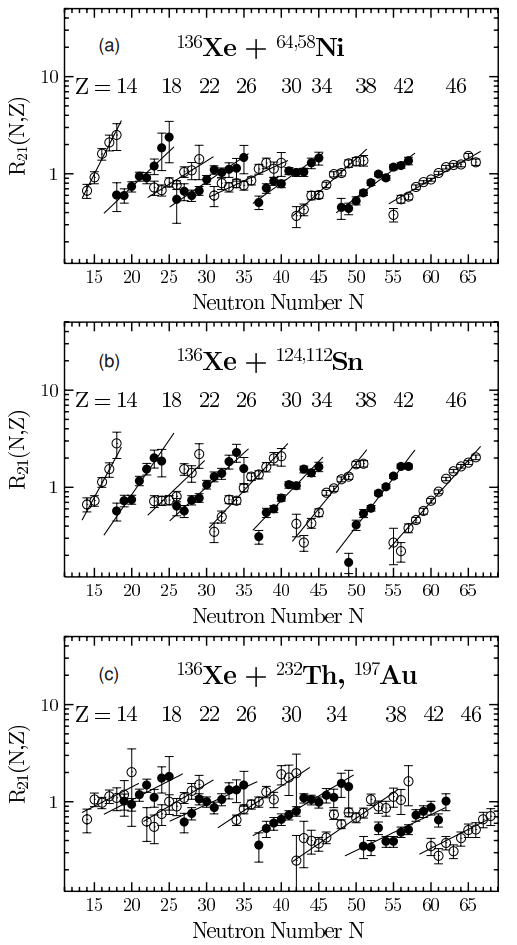}}
\caption{(Color online) Isoscaling of heavy residue fragments from (a) Xe~+~Ni, (b) Xe~+~Sn, and (c) Xe~+~Th/Au reactions.  The fragment yield ratio ($R_{21}$) is shown as a function of the neutron number for the labeled $Z$'s.  The lines represent exponential isoscaling fits to the data. Figure from Ref.~\cite{Soul06}  (Copyright 2009 by The American Physical Society).}
\label{f:residue}
\end{figure}

\par
A serious issue to be addressed is how to best extract information on the symmetry energy directly from the measured yield distributions such that the density and temperature dependence can be obtained. As discussed in the previous section, isoscaling of light or intermediate mass fragments can be accompanied by large secondary decay effects.  An alternative option that has been explored is the isoscaling of heavy residue fragment yields~\cite{Soul06,SOULI07,Ves04,Henz10}.  These yields may be less susceptible to the secondary decay effects in comparison to the light charged particles.  An example from the work of Souliotis \emph{et al.} of the heavy residue isoscaling is shown in Fig.~\ref{f:residue} for reactions around 25~MeV/nucleon~\cite{Soul06}.  At these energies, the reactions can be generally considered as deep-inelastic in which two heavy fragments are produced from the dissipative reaction along with light particle emission.  From the residue yields Souliotis \emph{et al.} where able to extract $C_{sym}$ as a function of the fragment excitation energy.  The results of the heavy residue and light fragment isoscaling are compared in Section~\ref{s:survey}.

\begin{figure}
\center
\resizebox{0.45\textwidth}{!}{\includegraphics{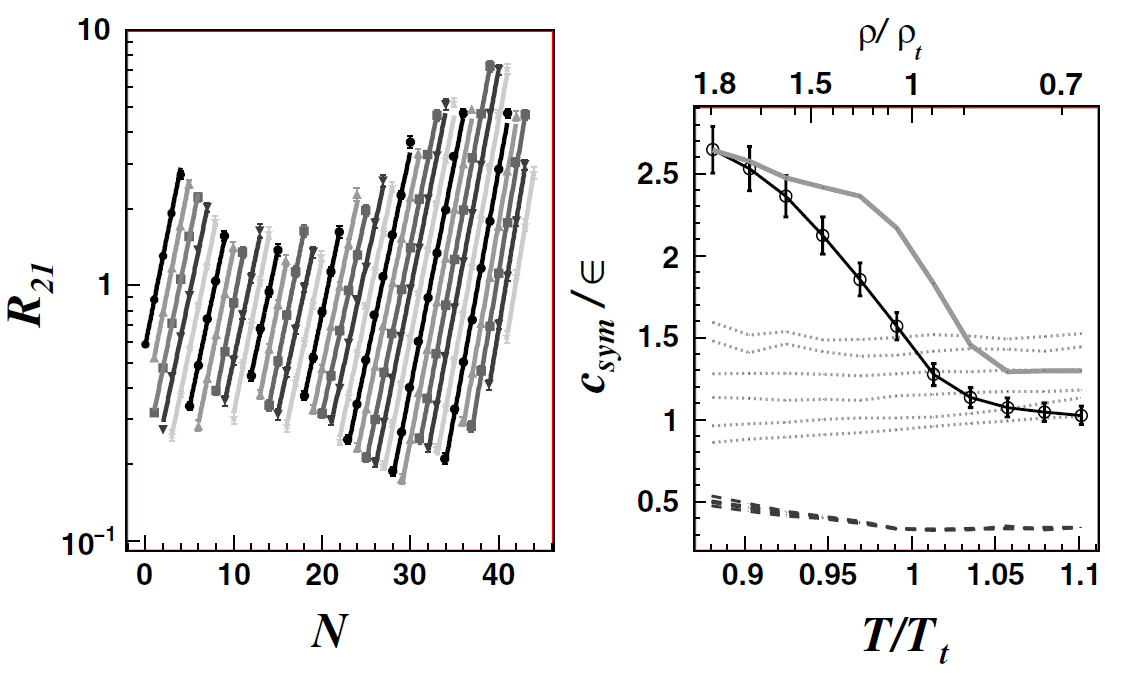}}
\caption{(Color online) Left panel: Isotopic fragment yield ratios from the lattice gas model.  Right panel: Reduced symmetry energy as a function of temperature from the lattice gas model.  The exact $T$-dependence of the symmetry energy is shown by the open circles. Results of isoscaling using $Z$~=~2~-~7 fragments with definitions from Eqs.~\ref{csym} and \ref{csym2} are shown as the black dashed lines and dotted grey lines, respectively.  The results using Eq.~\ref{csym2} with the heaviest fragment are shown as the thick solid grey line.  Figure from Ref.~\cite{Leh09}  (Copyright 2009 by The American Physical Society).}
\label{f:heavyfrag}
\end{figure}

Relevant to the heavy residue isoscaling is the theoretical work of Lehaut \textit{et al.} which, using the lattice gas model, suggested that the isoscaling of the largest fragment would provide a strong connection between the input and extracted symmetry energy coefficient~\cite{Leh09}.  As shown in Fig.~\ref{f:heavyfrag}, the open circles connected with a solid line represent the exact symmetry energy as a function of temperature for the simulation and the dashed black lines represent the symmetry energy extracted from isoscaling of the $Z$~=~2~-~7 fragments using the common source definition defined in Eq.~\ref{csym}.  The results show that the typical isoscaling analysis is unable to reproduce the correct trend or magnitude of the $T$-dependence of the symmetry energy.  Another option that was explored was using a different source definition such that,
\begin{equation}\label{csym2}
\alpha(Z) = \frac{4 C_{sym}(Z)}{T} \left[ \left( \frac{Z_{1}}{\langle A \rangle_{1}} \right)^{2} - \left( \frac{Z_{2}}{\langle A \rangle_{2}} \right)^{2} \right]
\end{equation}
where $\alpha$ and $C_{sym}$ are extracted for each $Z$ and $\langle A \rangle$ is the mean mass of the fragments for a given $Z$.  This definition is equivalent to the $\Delta_{liquid}$ definition used in Fig.~\ref{f:iso_compare}.  The dotted grey lines are the results using Eq.~\ref{csym2} for the $Z$~=~2~-~7 fragments and show a nearly flat $T$-dependence of the symmetry energy.  The weak relationship between the light fragment isoscaling results and the input symmetry energy was attributed to the domination of combinatorial effects in the lattice gas model. Despite the issues with the light fragments one of the most interesting aspects of the work was that Lehaut \textit{et al.} found that using the heaviest fragments from each event along with Eq.~\ref{csym2} the magnitude and $T$-dependence of the symmetry energy could be recovered relatively well (thick grey line).  The concept of heavy fragment isoscaling certainly warrants continued theoretical and experimental investigations as it could provide a clear link between experimental yields and the symmetry energy.

\subsection{Secondary decay corrections}

\par
While isoscaling of the heaviest fragment has been suggested to have a relatively direct link to $C_{sym}$, not all experimental setups and reaction mechanisms allow for the heaviest fragment to be identified and measured easily.  An alternative option is to measure the light fragment yields and use a theoretical model to account for the secondary decay effects~\cite{Bot02}. The use of the corrected light fragment yields can then provide information about the isoscaling of the primary fragment yields.  This, of course, is no longer a direct method since a theoretical model is required for the secondary decay correction.


\par
\begin{figure}
\center
\resizebox{0.34\textwidth}{!}{\includegraphics{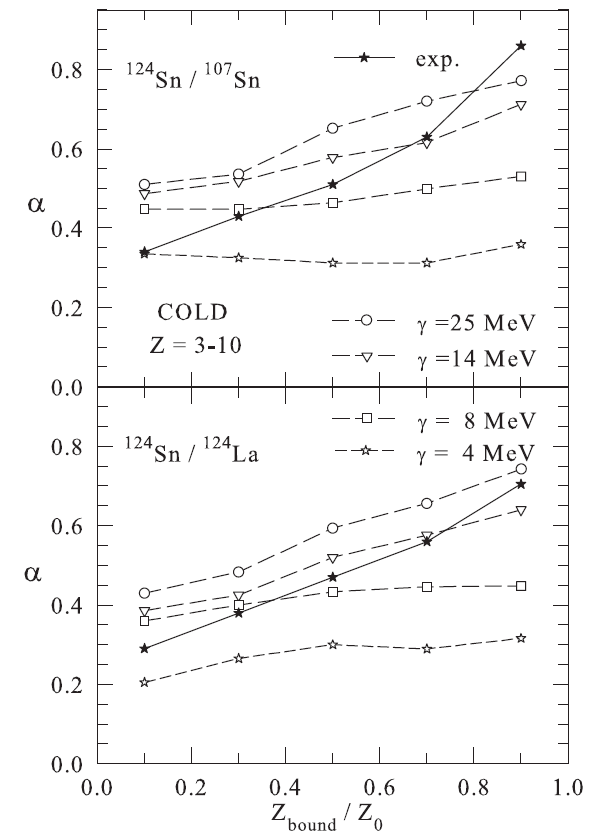}}
\caption{ Isoscaling parameter $\alpha$ from the $^{124}$Sn/$^{107}$Sn (top panel) and $^{124}$Sn/$^{124}$La (bottom panel) systems.  The solid triangles are from experimental $Z = 3-7$ isotopic yields and the open symbols represent results from the SMM model with different input values of the symmetry energy $\gamma$. Figure from Ref.~\cite{Ogul11}  (Copyright 2011 by The American Physical Society).}
\label{f:ogul}
\end{figure}

Ogul \textit{et al.} used this concept to extract information on $C_{sym}$ from a comparison of the experimental isoscaling parameter $\alpha$ with the Statistical Multifragmentation Model (SMM)~\cite{Ogul11}.  In Fig.~\ref{f:ogul} the experimentally extracted isoscaling parameters (solid stars) from relativistic fragmentation reactions are presented as a function of the ratio of the total measured bound charge to the projectile charge ($Z_{bound}$/$Z_{0}$), which is correlated to the impact parameter.  The isoscaling parameter decreases strongly with decreasing $Z_{bound}$/$Z_{0}$, which agrees well with the previously discussed results indicating a decrease in $\alpha$ with increasing $E^{*}$.  Instead of calculating $C_{sym}$ from the extracted $\alpha$, Ogul \emph{et al.} used the SMM model to simulate the decay of the excited sources with different input values of the symmetry energy (referred to as $\gamma$ in Fig.~\ref{f:ogul}).  The $\alpha$ parameter was extracted from the SMM model from the secondary decay products allowing for a direct comparison with the experimental data.  For each $Z_{bound}$/$Z_{0}$ value, the isoscaling parameter from SMM that best reproduced the experimental isoscaling parameter could be determined and, thus, the associated input $\gamma$ from the SMM model.  These values of $\gamma$ should then provide a direct link to the symmetry energy, free of any secondary decay effects.  For example, for the $^{124}$Sn/$^{124}$La system at $Z_{bound}$/$Z_{0}$~=~0.7 the experimental $\alpha$ is accurately reproduced by the SMM model using $\gamma = C_{sym} = 14$~MeV.  This provides an intermediate approach between the direct and indirect methods of probing the symmetry energy.

\subsection{Survey of isoscaling results}
\label{s:survey}

\begin{figure}
\center
\resizebox{0.44\textwidth}{!}{\includegraphics{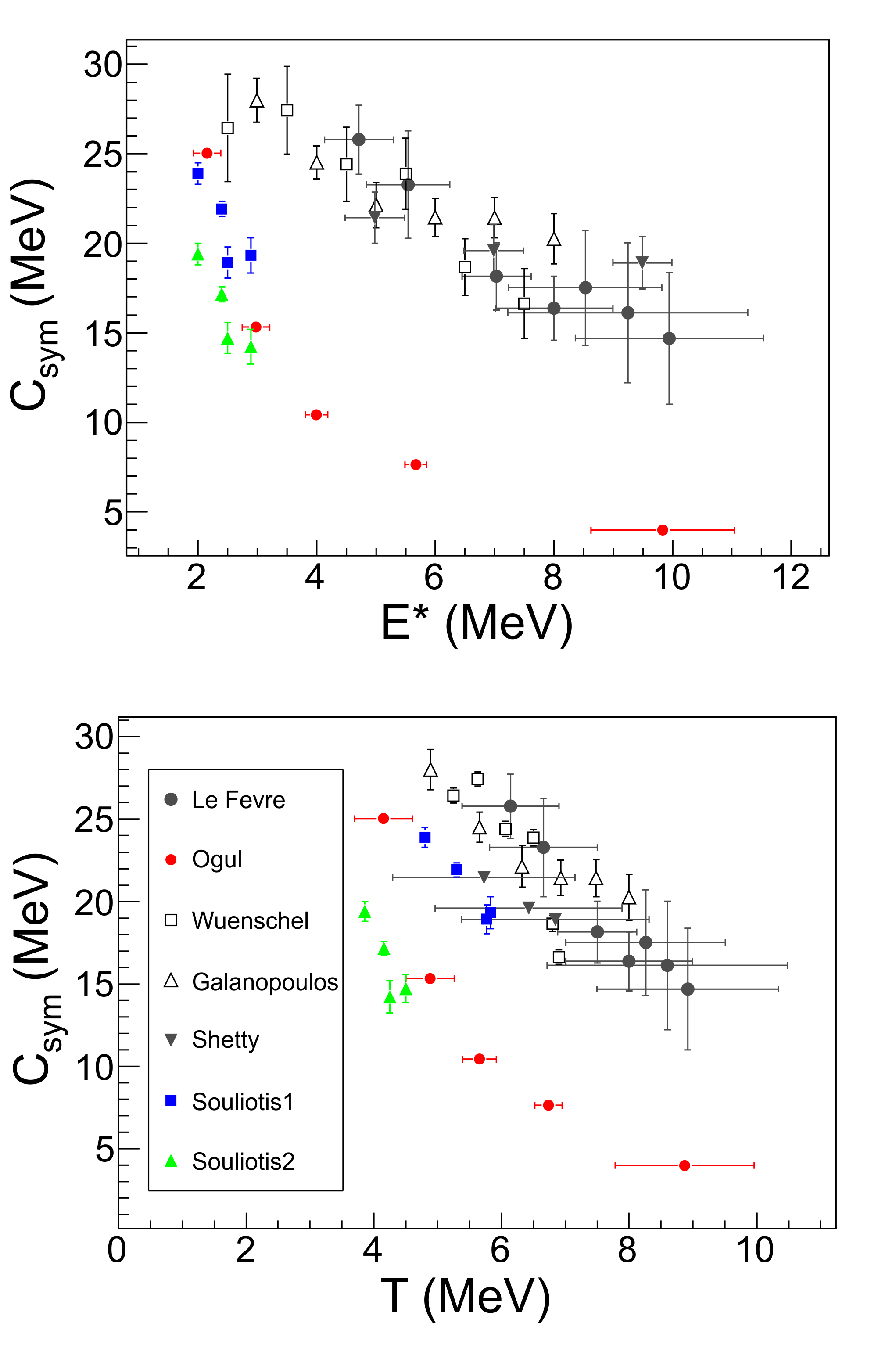}}
\caption{ Compilation of different experimental measurements of $C_{sym}$ from isoscaling analyses as a function of the excitation energy (top panel) and temperature (bottom panel). Pertinent details of each measurement are provided in Table~\ref{t:csym}. }
\label{f:csym}
\end{figure}

A survey of the experimental $E^{*}$ and $T$ dependence of the symmetry energy, extracted from isoscaling, was completed and is presented in Fig.~\ref{f:csym}.  Each measurement was performed differently in terms of the fragments used in the isoscaling, the method for extracting the temperature, and the calculation of the excitation energy of the source.  The details of each of the measurements are provided in Table~\ref{t:csym}.  In the case of measurements in which only the temperature or excitation energy was reported the corresponding excitation energy or temperature, respectively, was estimated using the simple Fermi Gas model relationship of $E^{*} = a T^{2}$, with $a = A/8$.  The values estimated using the Fermi Gas model are indicated in Table~\ref{t:csym}.

\par
In the top panel of Fig.~\ref{f:csym} the different results for the excitation energy dependence of the symmetry energy are presented and show two distinct groupings.  The results of Souliotis \textit{et al.} and Ogul \textit{et al.} indicate that the symmetry energy decreases rapidly with increasing $E^{*}$ over the range of 2-4~Mev.  The other isoscaling results present a consistently softer decrease of $C_{sym}$ with $E^{*}$.  In examining the details of the different isoscaling analyses in Table~\ref{t:csym} an important distinction is found between the two groupings.  The results of Ogul \textit{et al.} were effectively corrected for secondary decay using a simulation and Souliotis \textit{et al.} used the heaviest fragment from each event for the isoscaling.  As mentioned above, the isoscaling of the heaviest fragment was proposed as a promising observable for directly accessing the symmetry energy and may not require substantial secondary decay corrections.  Thus, the fact that the two analyses of Ogul \textit{et al.} and Souliotis \textit{et al.} agree is rather remarkable since both should be providing a more direct link to the symmetry energy since secondary decay effects should be minimized.    In comparison, the other isoscaling examples included light fragments in the analysis without secondary decay corrections and show a increased value of the symmetry energy.  These results qualitatively follow the expected trend from the multifragmentation models which indicate that the secondary decay increases the measured $C_{sym}^{out}$ for small input values of $C_{sym}^{in}$~\cite{Lef05,Mar13,Ogul11}.  It is worth noting that the agreement between the other analyses (using light fragments without secondary decay corrections) is impressive since different approaches were used in defining the source, measuring the temperature, and determining the excitation energy.  This consistency suggests that the isoscaling approach is relatively robust.

\begin{table*}
\begin{center}
\caption{Details about compilation of experimental isoscaling results.}
\begin{tabular}{c c c c c c}
\hline
\hline
Author                      &Reference            &Isoscaling Fragments    &Thermometer                  &$E^{*}$ calculation  &Sim-corrected decay\\
\hline
LeFevre     &\cite{Lef05}           &$Z=1-3$                 &Albergo $T_{HeLi}$              &Fermi gas  &No\\
Ogul       &\cite{Ogul11}           &$Z=3-10$        &Albergo $T_{HeLi}$, $T_{BeLi}$    &Fermi gas  &Yes\\
Wuenschel    &\cite{WUEN09THESIS,Wue09,Wue10}          &$Z=1-17$        &Momentum Fluctuations    &Calorimetry  &No\\
Galanopoulos    &\cite{Gal10}          &$Z=1-8$        &Fermi Gas    &Calorimetry  &No\\
Shetty    &\cite{Shetty07}          &$Z=3-7$        &SMM simulation    &BNV simulation  &No\\
Souliotis1    &\cite{Soul06,SOULI07}          &Heavy residue        &Fermi gas    &Binary kinematics  &No\\
Souliotis2    &\cite{Soul06,SOULI07}          &Heavy residue        &Expanding mononucleus    &Binary kinematics  &No\\
\hline
\hline
\end{tabular}
\label{t:csym}
\end{center}
\end{table*}

In the bottom panel of Fig.~\ref{f:csym}, the same survey of data is presented as a function of temperature.  The results present a similar picture to the excitation energy dependence of the symmetry energy.  One difference is that the Souliotis1 and Souliotis2 results are now separated.  The reason for the difference is based on how $T$ was derived.  In the experimental measurement of Souliotis \emph{et al.} the $E^{*}$ of the heavy residue is calculated from the kinematics of the reaction and then converted into $T$.  The conversion of $E^{*}$ into $T$ was done using the Fermi Gas model (Souliotis1) and using the expanding mononucleus model (Souliotis2).  The different relationships between $E^{*}$ and $T$ lead to different values of $C_{sym}$.  The filled blue square represent the results in which the temperature was calculated using the Fermi gas model and the filled green triangles represent the use of the expanding mononucleus model.  This complicates the results since the agreement of the Souliotis \textit{et al.} isoscaling with either the Ogul \textit{et al.} or other experiments strongly depends on the choice of the temperature calculation.  However, the expanding mononucleus model should provide a better description of the heavy residues produced from the deep-inelastic reactions.  In either case, the results of Ogul \textit{et al.}, which are simulation dependent, suggest that extracted $C_{sym}$ from the other analyses is systematically larger.

\par
Overall, the results of the isoscaling survey demonstrate the power of the isoscaling approach for accessing information on the symmetry energy and the critical need for continued experimental measurements.  In particular, we highlight the interesting results of Souliotis \textit{et al.} using heavy fragment isoscaling as a possible avenue for accessing the symmetry energy without requiring significant secondary decay corrections~\cite{Soul06,SOULI07}.  However, in coincidence with measuring the heavy fragments, it is important to measure reaction products that can provide insight into the temperature, excitation energy, and possibly the density of the system.  As shown by Shetty \textit{et al.}, the strong decrease in the symmetry energy as a function of temperature and excitation energy appears to be largely dominated by the decreasing density of the source~\cite{Shetty07}.  Ideally, an experiment in which all of these variables could be extracted would demonstrate significant progress towards understanding the temperature, excitation energy, and density dependence of the symmetry energy.

\section{Indirect Measurements}
  \label{sec:1}

\par
Comparison of experimental data and transport calculations allows for the evaluation of the nuclear EoS through the effective nucleon-nucleon interaction implemented in the simulation~\cite{LI01,Li08,Tsang12}.  In the heavy-ion collisions the density, temperature, and pressure of the system will evolve with time.  The transport calculations attempt to simulate this evolution.  Thus, if a given nucleon-nucleon interaction shows improved agreement with experimental data then the density and temperature dependence of the symmetry energy can be inferred from the interaction.

\subsection{Comparison of experiment and transport theory}

\par
Flow observables from heavy-ion collisions have been repeatedly used to examine the nuclear EoS~\cite{REISDORF97,HERR99,OGILVIE90,CUSSOL02,KROF92,MAGE00,WESTFALL93,LI98,MAGE002,DITORO06,PAK97FLOW,PAK97EBAL,SCALONE99}.  Experimentally the transverse flow is often quantified as the slope of the average in-plane momentum, $\left\langle p_{x} \right\rangle $, over the mid-rapidity region.  The flow parameter (F) is then defined as,
\begin{equation}
\label{eqF}
F = \frac{\partial \left\langle p_{x} \right\rangle}{\partial Y_{r}} \lvert_{Y_{r}=0}
\end{equation}
where Y$_{r}$ represents the reduced rapidity, which is equal to the center-of-mass rapidity of the fragment scaled by the center-of-mass rapidity of the projectile \linebreak (Y$_{r}$=Y$_{cm,frag}$/Y$_{cm,proj}$).  In the Fermi energy regime the rapidity, Y, is nearly equivalent to the parallel, or beam direction, velocity (v$_{z}$). The flow provides information on the collective movement of nuclear matter in mid-peripheral collisions which should be strongly dependent on the nuclear potential and nucleon-nucleon collisions~\cite{Danielewicz2002,Das93}.  Thus, flow can be used to probe the nuclear EoS but requires a simulation of the heavy-ion collision dynamics.   The contribution of Russotto and Trautmann to this EPJA issue discusses the use of flow measurements at relativistic energies to probe the nuclear EoS~\cite{RUSSOTTO11}.

\begin{figure}
\center
\resizebox{0.39\textwidth}{!}{\includegraphics{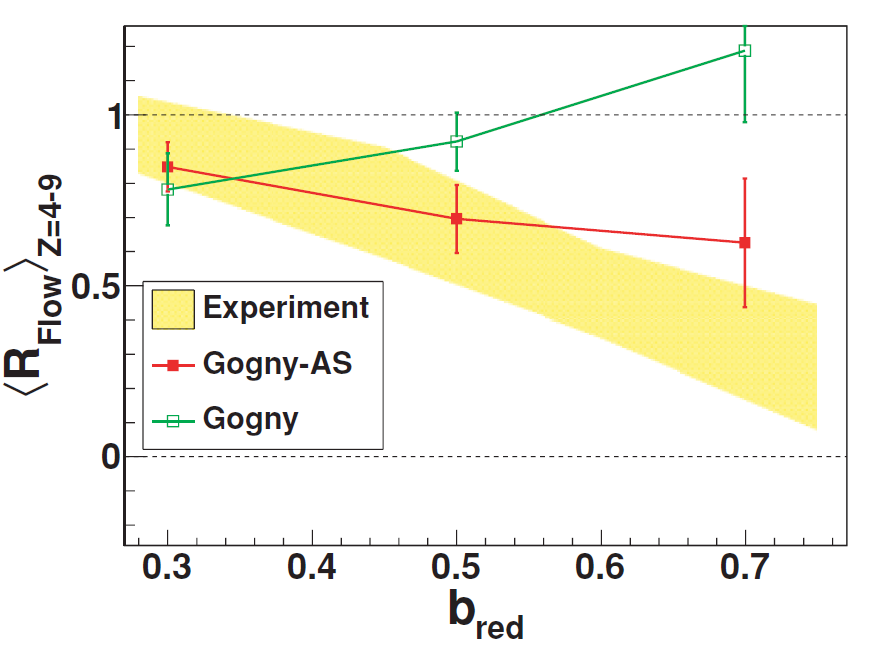}}
\caption{(Color online) $R_{Flow}$ ratio, as defined in Eq.~\ref{eqFlowRatio}, is shown as a function of the reduced impact parameter ($b_{red}$)) from experimental and simulated data.  The AMD simulation is shown using two different interactions: Gogny and Gogny-AS.  Figure from Ref.~\cite{Koh10}  (Copyright 2010 by The American Physical Society).}
\label{f:Koh10}
\end{figure}

\par
In Fig.~\ref{f:Koh10} the ratio of the intermediate mass fragment flow (IMF) from three different systems is shown as a function of the reduced impact parameter $b_{red}$~\cite{Koh10}.  The ratio, $R_{Flow}$, was defined as,
\begin{equation}
\label{eqFlowRatio}
R_{Flow} = \dfrac{\overline{\left\langle p_{x}/A \right\rangle}_{^{64}Zn} - \overline{\left\langle p_{x}/A \right\rangle}_{^{70}Zn} }  {\overline{\left\langle p_{x}/A \right\rangle}_{^{64}Ni} - \overline{\left\langle p_{x}/A \right\rangle}_{^{70}Zn} }
\end{equation}
where $\left\langle p_{x}/A \right\rangle$ is the average transverse flow from the $^{64}$Zn+$^{64}$Zn, $^{70}$Zn+$^{70}$Zn, and $^{64}$Ni+$^{64}$Ni systems.  The $\left\langle p_{x}/A \right\rangle$ was quantified as the average in-plane transverse momentum from $0.0 \leq Y_{r} \leq 0.45$ since detector thresholds produced incomplete IMF detection at negative rapidities~\cite{Koh10}.  The experimental data demonstrates a strong decrease of $R_{Flow}$ with increasing impact parameter.  Simulations of the reaction systems from the antisymmetrized molecular dynamics model (AMD) using two different nucleon-nucleon interactions (Gogny and Gogny-AS) were compared to the experiment and are shown in Fig.~\ref{f:Koh10}.  A strong difference in the impact parameter dependence of $R_{Flow}$ is observed between the different interactions used in the AMD model.  The Gogny-AS interaction is able to reproduce the experimental trend better than the Gogny interaction~\cite{Koh10}.

\begin{figure*}
\center
\resizebox{0.66\textwidth}{!}{\includegraphics{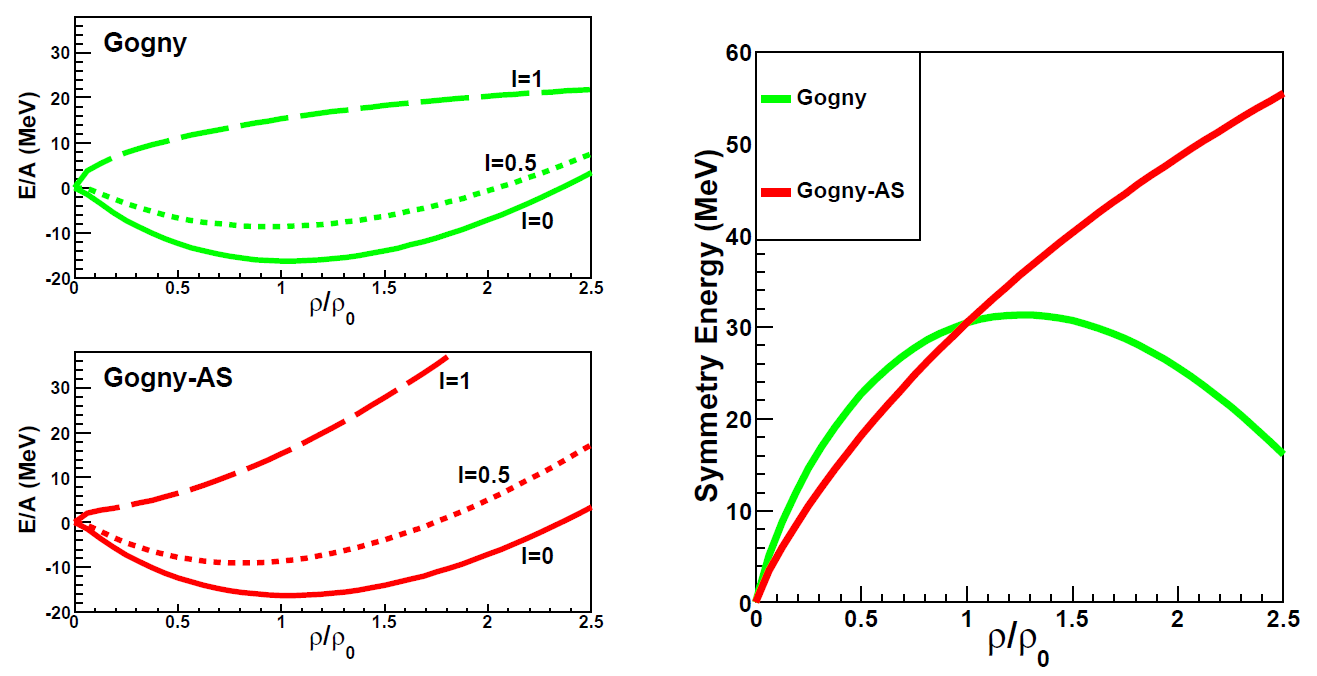}}
\caption{(Color online) (Left panels) Binding energy per nucleon as a function of density for infinite nuclear matter with different asymmetries ($I = N-Z/A$) calculated using the Gogny and Gogny-AS interactions.  (Right panel) The density dependence of the symmetry energy calculated from the difference in the equation of state of symmetric nuclear matter and pure neutron matter.}
\label{f:gogny}
\end{figure*}

\par
As discussed above, this type of indirect analysis then allows for information on the symmetry energy to be inferred from the interaction that best reproduces the experimental data.  The nuclear equations of state calculated using the Gogny and Gogny-AS interactions are shown in Fig.~\ref{f:gogny}~\cite{ONO_04,OnoAMD02}.  The panels on the left show the binding energy of nuclear matter with different asymmetries ($I$) as a function of density and the panel on the right shows the corresponding density dependence of the symmetry energy for each interaction.  From the comparison of the experiment and AMD simulation, we can infer that the density dependence of the symmetry energy produced with the Gogny-AS interaction is more accurate than that of the Gogny interaction.  Similarly, one could extract information on the temperature dependence of the symmetry energy using the interactions within the correct formalism.  Thus, with the indirect approach it is not necessary to measure the exact density or temperature of the reaction source since the entire evolution of the system is naturally taken into account.  However, it is useful to understand what are the regions of density and temperature that the extracted information would have the most fidelity.  For example, in the flow analysis discussed above it was determined that the observable was sensitive to densities below the saturation density~\cite{Koh10}.

\begin{figure}
\center
\resizebox{0.39\textwidth}{!}{\includegraphics{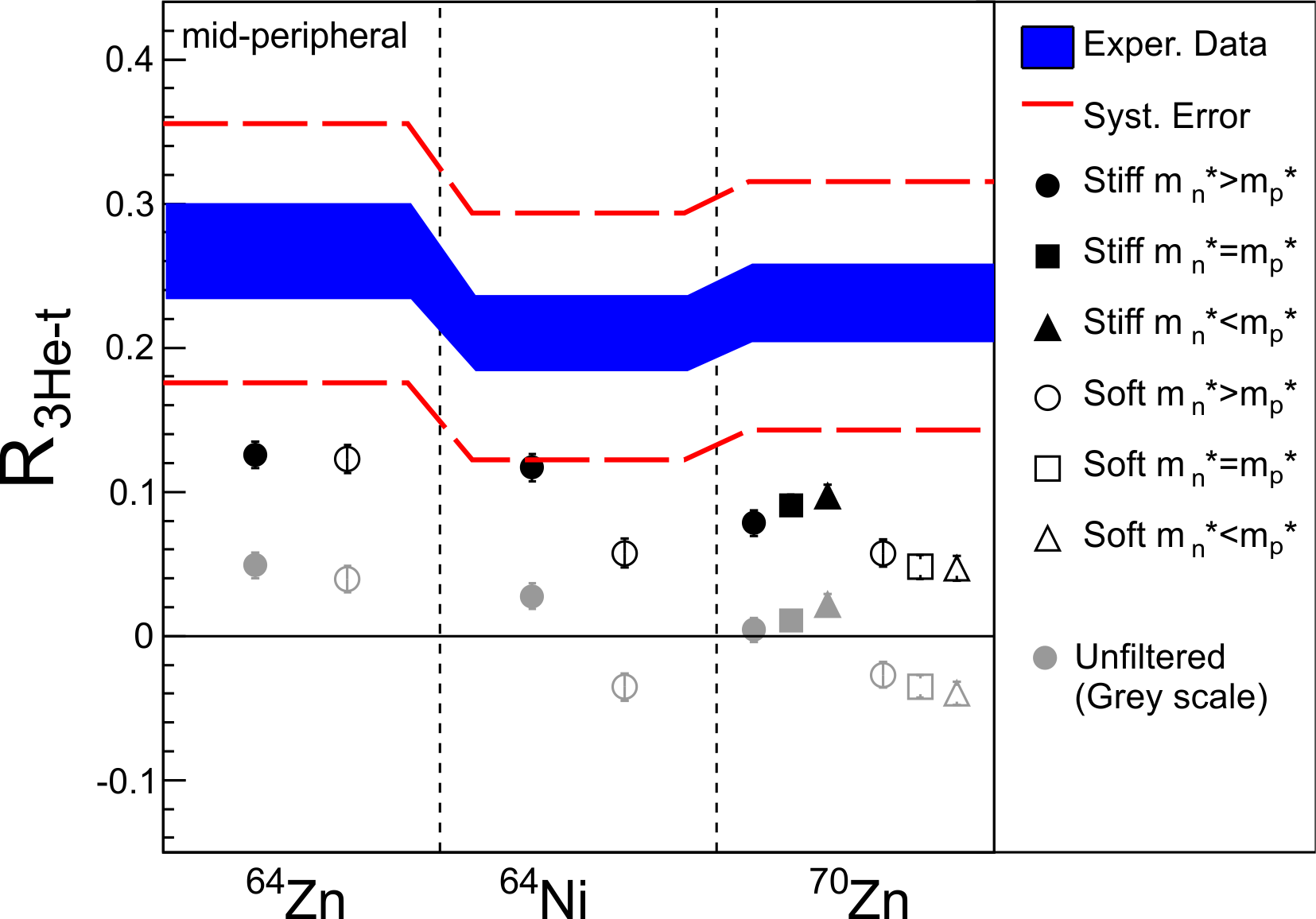}}
\caption{(Color online) Ratio of the triton and $^{3}$He flow ($R_{3He-t}$) from the symmetric $^{64}$Zn, $^{64}$Ni, and $^{70}$Zn systems. Experimental measurement is shown as the solid band with the long dashed red lines representing the limits of the systematic errors. The results from the SMF model with different forms of the symmetry energy and effective mass ratios are shown by the markers. }
\label{f:IsoFlow}
\end{figure}

In Fig.~\ref{f:IsoFlow} the experimentally measured triton to $^{3}$He flow ratio is shown from symmetric heavy-ion collisions of Zn and Ni nuclei.  The ratio was defined as $R_{3He-t} = (F^{3He} - F^{t}) / (F^{3He} + F^{t})$, where $F^{3He}$ ($F^{t}$) is the transverse flow of the $^{3}$He (triton) fragments~\cite{KOHLEYLCP}.  The experimental data is compared to the Stochastic Mean-Field model (SMF) which uses the BGBD interaction~\cite{BARAN05,RIZZO04,GIORDANO10}.  Besides varying the density dependence of the symmetry energy (labeled as stiff or soft in the legend), the relative effective mass ($m^{*}$) of the neutrons and protons was varied for the $^{70}$Zn system.  Changing the relative effective mass is equivalent to changing the isospin dependence of the momentum dependent part of the interaction within the transport calculation, but the relative effective mass itself is an important parameter.  While the simulation was unable to fully reproduce the experimental data, the results demonstrate a reasonable sensitivity of the light fragment flows to the density dependence of the symmetry energy and a weak sensitivity to the neutron-proton effective mass splitting for low energy reactions (~35 MeV/nucleon).  The results of the SMF model provide motivation for future experimental and theoretical light fragment flow investigation.

\par
\begin{figure}
\center
\resizebox{0.42\textwidth}{!}{\includegraphics{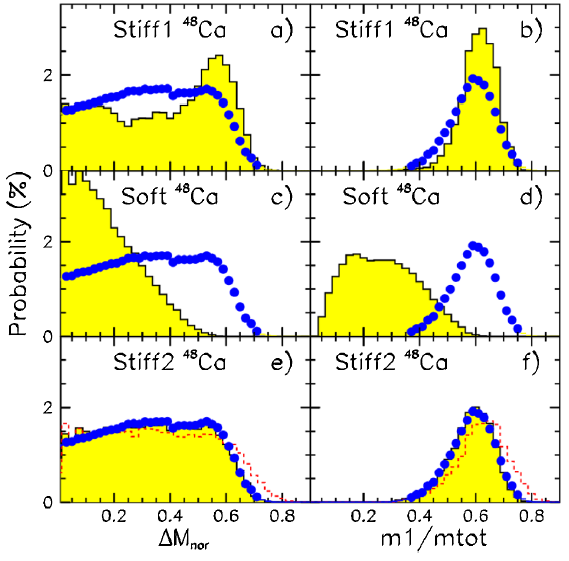}}
\caption{(Color online) The probability distribution for $\Delta M_{nor}$ and $m1/mtot$ for events from the Ca+Ca heavy-ion collisions.  The experimental data is shown as the solid blue circles and the CoMD simulations as the histograms. Figure from Ref.~\cite{Amo09}  (Copyright 2009 by The American Physical Society).}
\label{f:chimera}
\end{figure}

While flow observables have been commonly discussed in reference to exploring the symmetry energy many other dynamical observables should be sensitive to the isospin dependent part of the nucleon-nucleon interaction. For example, the CHIMERA collaboration examined low-energy incomplete fusion reactions of Ca+Ca and compared the results with the constrained molecular dynamics simulation (CoMD)~\cite{Amo09,Card12}. The two heaviest fragments were identified in each event ($m1$ and $m2$).  Two observables $\Delta M_{nor} = (m1-m2)/mtot$ and $m1/mtot$ were constructed from the experimental and simulated data, where $mtot$ is the total mass of the entrance channel.  A comparison of these observables from the simulation (solid histograms) and experiment (solid circles) are shown in Fig.~\ref{f:chimera}.  The Skyrme interaction used in the CoMD model was varied to produce different forms of the symmetry energy. Approximating the form factor of the symmetry potential as $(\rho / \rho_{0})^{\gamma}$, the stiff1, stiff2, and soft forms of the symmetry energy correspond to $\gamma$~=~1.5, 1.0, and 0.5, respectively~\cite{Card12}.   The results show that the dynamical break up or incomplete fusion of the Ca+Ca systems is strongly connected to the isospin dependent part of the Skyrme interaction and that the stiff2 density dependence of the symmetry energy greatly improves the agreement with the experimental data relative to the other forms of the symmetry energy.

It is important to recognize two of the seminal works that pushed the concept of using transport theory to constrain the density dependence of the symmetry energy through comparison with heavy-ion collisions.  In 2004, both Tsang \emph{et al.}~\cite{TSANG04} and Shetty \emph{et al.}~\cite{Shetty04} published works comparing measured heavy-ion collision observables to transport calculations in which the form of the symmetry energy had been varied.  Specific constraints on the symmetry energy were not reported in either case but extreme forms of $E_{sym}(\rho)$ were able to be discarded.  About 3 years later, Shetty \emph{et al.} reported one of the first constraints derived from heavy-ion collisions through comparison of the isoscaling parameter with the AMD model~\cite{Shetty07,Shetty_07}.  The reported constraint of $E_{sym} = 31.6 (\rho / \rho_{0})^{\gamma}$ with $\gamma = 0.6 - 1.05$ is still in good agreement with more recent constraints~\cite{Shetty_07} (see discussion in Section~\ref{constraints}).

\begin{figure}
\center
\resizebox{0.49\textwidth}{!}{\includegraphics{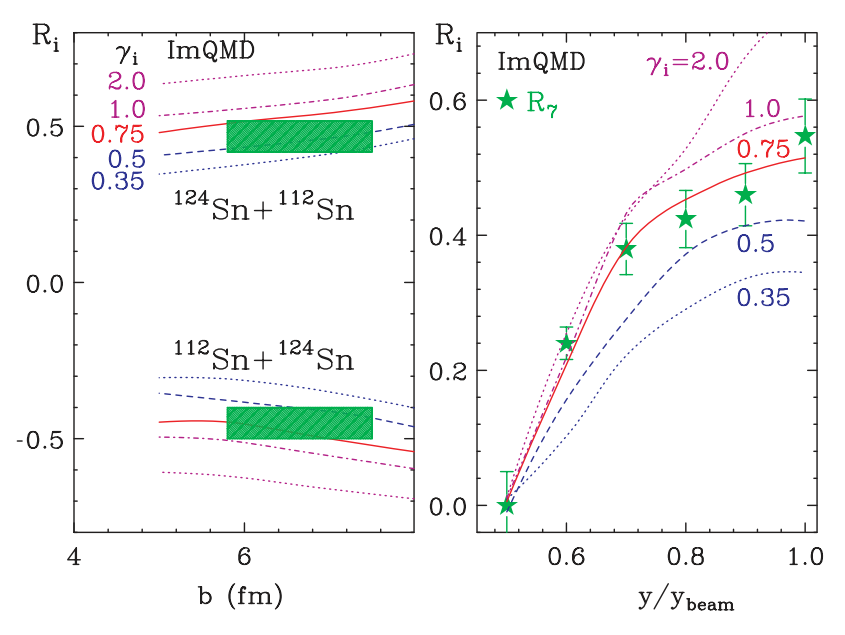}}
\caption{(Color online) (Left panel) Isospin transport ratio ($R_{i}$) calculated from the isoscaling parameter $\alpha$ as a function of the impact parameter from Sn+Sn reactions.  (Right panel) Isospin transport ratio calculated from $A = 7$ yield ratios as a function of the fragment rapidity.  The experimental data is shown as the green bars (left) and green solid triangles (right).  The ImQMD simulation results are shown as the solid and dashed lines representing calculations with different values of $\gamma$.  Figure from Ref.~\cite{TSANG09}  (Copyright 2009 by The American Physical Society).}
\label{f:Tsang}
\end{figure}

The isospin transport ratio ($R_{i}$) observable measured by Tsang \emph{et al.} provides a measure of the isospin equilibration that occurs in the heavy-ion collisions.   The initial measurement in Sn~+~Sn reactions inspired a number of theoretical works which demonstrated a strong sensitivity of the observable to the density dependence of the symmetry energy~\cite{TSANG04,Chen_05,Steiner_05,LI05,Baran_05}.  More restrictive constraints on the symmetry energy were obtained from the two isospin diffusion measurements presented in Fig.~\ref{f:Tsang}.  In the left panel the isospin transport ratio is shown from both the experiment (solid green bars) and a simulation using the improved quantum molecular dynamics model (ImQMD).  The density dependence of the symmetry energy was varied within the ImQMD simulation and is represented by the different values of $\gamma$.  The isospin transport ratio was calculated using the isoscaling parameters, $\alpha$, determined from the fragment emission from different Sn~+~Sn systems.  Thus, rather than using the isoscaling method to attempt to directly probe the EoS, it was used instead as a observable sensitive to the isospin dependent part of the nucleon-nucleon interaction.  Similarly, the right panel shows the isospin transport ratio calculated using the yield ratios of $A = 7$ fragments as a function of the fragment rapidity.  In both cases, the results show a sensitivity to the $\gamma$, or symmetry energy, used within the ImQMD simulation and allowed for constraints on the form of the EoS to be obtained from the comparison.  These extracted constrains (shown in Section~\ref{constraints}) have significantly impacted both the nuclear and astrophysics communities.

\par
Of course it is difficult to present all the measured and/or predicted observables that heavy-ion collisions offer for accessing information about the nuclear EoS.  Other promising observables for probing the EoS through comparisons with transport calculations include neutron-to-proton ratios~\cite{FAMIANO06,ZHANG08}, neutron enrichment of mid-velocity fragments in semi-peripheral collisions~\cite{KOHLEYLCP,DeFil12,DITORO06NECK}, light fragment ratios (such a $t$ to $^{3}He$)~\cite{SCALONE99,CHEN03,YONG09}, and $\pi^{-}/\pi^{+}$ ratios (applicable at higher energies)~\cite{XIAO09,Xie13,FENG10,Gao12,Xu13,Hong13}. The complex and dynamic nature of heavy-ion collisions over different energies and impact parameters creates a wide variety of situations in which nuclear matter can be created at varying temperatures, densities, and pressures.

\subsection{Constraints on $E_{sym}(\rho)$ from heavy-ion collisions}
\label{constraints}

\begin{figure}
\center
\resizebox{0.40\textwidth}{!}{\includegraphics{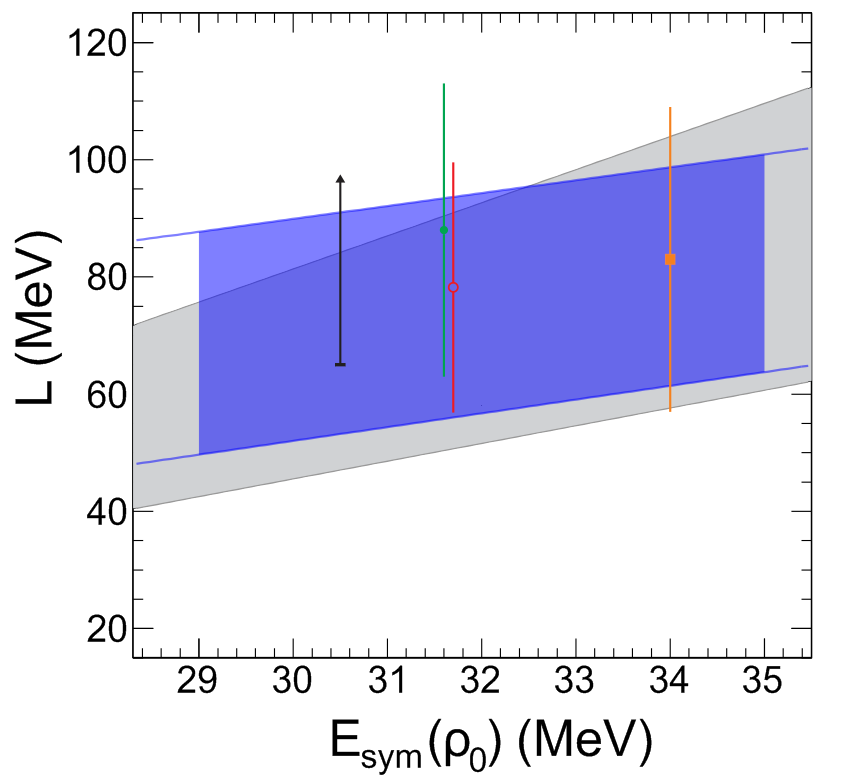}}
\caption{(Color online) The heavy-ion collisions constraints on the slope ($L$) and magnitude at $\rho_{0}$ of the symmetry energy from Refs.~\cite{TSANG09} and \cite{Kohley13} are shown as the solid grey and shaded blue regions, respectively.  Additional constraints at given values of $E_{sym}(\rho_{0})$ from heavy-ion collisions are shown from Refs.~\cite{Koh10}, \cite{Li08}, \cite{Shetty_07}, and~\cite{RUSSOTTO11} as (left to right) black lower-limit, solid green circle, open red circle, and solid orange square, respectively.  The $E_{sym}(\rho_{0})$ of the red square was offset by 0.1~MeV for clarity. }
\label{f:Lesym}
\end{figure}

\par
Much like the survey of the isoscaling analyses presented in Fig.~\ref{f:csym}, a compilation of the comparisons of experimental heavy-ion reaction measurements and transport calculations can provide insight into the current constraints and understanding of the asymmetry dependence of the nuclear EoS.  A common approach for comparing the different constraints is to compare the Taylor expansion coefficients of the density dependence of the symmetry energy around the saturation density:
\begin{subequations}
\begin{align}
\label{eqTaylor}
&E_{sym}(\rho) = E_{sym}(\rho_{0}) + L\kappa + \frac{K_{sym}}{2} \kappa^{2} + O(\kappa^{3})\\
&L = 3 \rho_{0} \frac{d E_{sym}(\rho)}{d \rho}\mid_{\rho = \rho_{0}} \\
&K_{sym} = 9 \rho_{0}^{2} \frac{d^{2} E_{sym}(\rho)}{d \rho^{2}}|_{\rho = \rho_{0}}
\end{align}
\end{subequations}
where $\kappa = (\rho - \rho_{0})/3 \rho_{0}$.  In Fig.~\ref{f:Lesym} the current constraints extracted from heavy-ion collisions are presented in a $2D$ plot showing the slope ($L$) as a function of the magnitude of the symmetry energy at saturation density [$E_{sym}(\rho_{0})$].

The two-dimensional constraint shown in grey was extracted from the isospin diffusion work of Tsang \emph{et al.}~\cite{TSANG09} and has become a benchmark comparison for studies of the nuclear EoS from heavy-ion collisions to neutron star observations~\cite{Tsang12}.  Recently, Kohley \emph{et al.} used radioactive-ion beam induced reactions to derive constraints on the form of the symmetry energy which are presented in Fig.~\ref{f:Lesym} as the shaded blue region~\cite{Kohley13}.  Good agreement is observed between the two constraints even though the reaction mechanism, observable, and transport theory was different.  Additional constraints on $L$ at specific values of $E_{sym}(\rho_{0})$ from heavy-ion collisions are also presented.  Around a value of $E_{sym}(\rho_{0}) = 32.5$~MeV (the value based on fitting binding energies~\cite{Moller12}) the results suggest the slope of the symmetry energy to be near 70~MeV.  While it is clear that further experiments and theoretical comparisons are necessary to continue to define the form of the symmetry energy it is encouraging to see relatively consistent results coming from very different analyses.  It is worth noting that higher order Taylor expansion terms, such as the curvature ($K_{sym}$), will also be important to compare for constraints farther away from the saturation density.

\begin{figure*}
\center
\resizebox{0.65\textwidth}{!}{\includegraphics{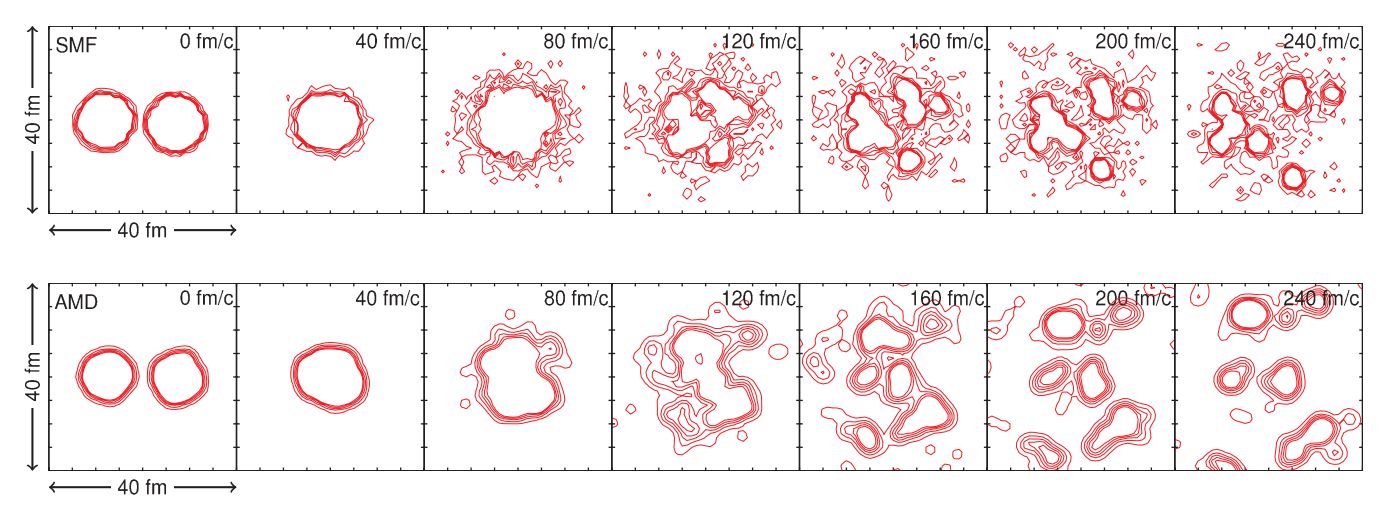}}
\caption{(Color online) Contours of the density projected on to the reaction plane from the SMF and AMD models of a $^{112}$Sn + $^{112}$Sn collision at 50 MeV/nucleon. Figure taken from Ref.~\cite{COLONNA10}  (Copyright 2010 by The American Physical Society). }
\label{f:maria}
\end{figure*}

\subsection{Transport theory discussion}

\par
While the constraints show some consistency, they are strongly dependent on comparisons with theoretical transport calculations.  Therefore, the results depend not only on the form of the symmetry energy but also on the treatment of the nuclear dynamics in that specific model~\cite{ONO06}.  We would like to emphasize the importance of the work by Rizzo \emph{et al}.~\cite{RIZZO07} and Colonna \emph{et al}.~\cite{COLONNA10} in which detailed studies of different transport calculations are examined.  The comparison of the SMF and AMD models demonstrated how the description of the nuclear dynamics can change the multi-fragmentation process and, thus, effect the HIC observables used in constraining the symmetry energy.  Fig.~\ref{f:maria} shows a comparison of a $^{112}$Sn + $^{112}$Sn reaction at 50 MeV/u from the AMD and SMF models.  From the density contours alone differences in the dynamics can be observed.  The AMD model shows a faster expansion after the initial collision while the SMF model presents a larger nucleon emission (or an increased "gas" phase).  The results of the study showed that the one-body (SMF) versus many-body (AMD) effects can significantly change the fragmentation properties of the reactions. One observable suggested to be sensitive to these correlations is the production of light ($Z > 2$) and heavy ($Z > 6$) intermediate mass fragments.

\par
It is clear that accurate constraints on the density dependence of the symmetry energy will require consistent theoretical descriptions of the experimental data.  For example, examination of experimental $\pi^{-}$/$\pi^{+}$ ratios with the mean-field model IBUU04 suggested a soft symmetry energy at high densities~\cite{XIAO09}.  In contrast, a stiff form of the symmetry energy was found to better reproduce the same $\pi^{+}$/$\pi^{-}$ data when using the improved isospin dependent quantum molecular dynamics (ImIQMD) model~\cite{FENG10}.  Similarly, the sensitivity of neutron-to-proton (\textit{n}/\textit{p}) ratios to the density dependence of the symmetry was shown to be widely different depending on the choice of the theoretical simulation~\cite{FAMIANO06,ZHANG08}.  Kohley \emph{et al.} also compared the $R_{Flow}$ observable (shown in Fig.~\ref{f:Koh10}) to three different transport calculations to examine the dependence of the sensitivity of the symmetry energy to the description of the reaction dynamics~\cite{Koh12}.  Overall, the different studies demonstrate the importance in comparing multiple theoretical calculations to the same HIC observable(s) in order to validate any constraints.  Along with comparison between models, it is also crucial to understand how the different parameters within a single model can effect the outcome.  Zhang \emph{et al.}~\cite{Zhang12} and Coupland \emph{et al.}~\cite{Coup11} explored how adjustments to the input physics (in-medium cross section and momentum dependent interactions) of the transport calculations can also affect the resulting HIC observables.


\section{Summary}
Production of nuclear matter at densities, temperatures, and pressures away from ground state nuclear matter is vital for mapping out the nuclear EoS.  Heavy-ion collisions provide, likely, the best mechanism for producing and studying nuclear matter as a function of density, temperature, and pressure in a terrestrial laboratory.  In the presented review two approaches in using heavy-ion collisions to probe the EoS were presented: (1) the direct approach in which the measured fragment yields are linked ``directly'' to the symmetry energy and (2) the indirect approach based on the comparison of transport calculations with experimentally measured observables to identify the effective nucleon-nucleon interactions (and corresponding EoS) which best reproduce the data.

While much effort has been focused on extracting the density dependence of the symmetry energy it is important to recognize that for many of these studies the temperature is not constant.  For measurements based on ground state binding energies and nuclear structure the temperature is well determined.  However in heavy ion collisions the temperature, as well as the density, changes through the course of the collision.  The temperature dependence of the symmetry is an important parameter in understanding the dynamical evolution of massive stars.

Techniques have been developed which can allow for information on the density, temperature, and symmetry energy to be extracted, with assumptions, directly from the heavy-ion collisions measurements.  A compilation of the results from these measurements was shown in Fig.~\ref{f:csym} for the extracted $C_{sym}$ as a function of excitation energy and temperature.  In the cases that the analysis was similar the results show that the isoscaling method is relatively robust producing results in good agreement.  Of particular interest was the results of Ogul which were corrected for secondary decay effects and the heavy residue isoscaling results of Souliotis.  The results agreed well and suggest that the heavy residue isoscaling could be a probe that is less susceptible to secondary decay effects.  Further investigation of heavy fragment isoscaling is certainty encouraged.  In order to move forward in the direct measurements more exclusive experiments must be done that measure in coincidence both the density and the temperature as well as the observable connected to the symmetry energy.

The indirect approach is strongly dependent on the theoretical transport calculations but has been demonstrated to provide a powerful avenue for constraining the density dependence of the symmetry energy.  The experimental and theoretical efforts in studying the isospin diffusion process produced competitive constraints on the slope and magnitude of the symmetry energy.  Further verification of these constraints was provided through a separate experiment using a radioactive ion beam induced reaction and extracted similar constraints.  The current constraints on the density dependence of the symmetry energy were presented in Fig.~\ref{f:Lesym}.  While rather large uncertainty remains, the agreement between the different constraints is impressive since they derive from different reaction mechanisms, observables, and transport theories.  Further development of the models and comparison of various models with multiple experimental observables at the same time will be important for continued refinement of the heavy-ion collision constraints.

The current progress and results presented in this review should demonstrate the immense opportunities available in the field for continuing to constrain the nuclear EoS with heavy-ion collisions.

\begin{acknowledgement}
This work is supported by the National Science Foundation under Grant No. PHY-1102511, the Robert A. Welch Foundation under grant A-1266, and the Department of Energy under grant  DE-FG03-93ER-40773.
\end{acknowledgement}


\end{document}